\documentclass{aastex63}

\usepackage{newtxtext,newtxmath}
\usepackage[T1]{fontenc}
\usepackage{ae,aecompl}

\usepackage{graphicx}	% Including figure files
\usepackage{amsmath}	% Advanced maths commands

\usepackage{amssymb}	% Extra maths symbols
\usepackage{natbib}
\usepackage{xfrac}

\newcommand{\appropto}{\mathrel{\vcenter{
  \offinterlineskip\halign{\hfil$##$\cr
    \propto\cr\noalign{\kern2pt}\sim\cr\noalign{\kern-2pt}}}}}
\received{xxx}
\revised{yyy}
\accepted{zzz}
\shorttitle{On the Evidence for Violation of the Equivalence Principle in Disk Galaxies}
\shortauthors{Sargent et al.~2025}

\begin{document}

\title{On the Evidence for Violation of the Equivalence Principle in Disk Galaxies}
%\TitleCitation{On the Violation of the Equivalence Principle}

% Authors, for the paper (add full first names)
\author{Corey Sargent}
\thanks{Deceased. This publication is dedicated to the memory of Dr. Corey Sargent, who tragically passed away on11 June 2022, at the age of 25, and to his parents, Tracy Sargent and Gerard Carelli. At the time of his passing, Corey was developing the model at the heart of this article. To his fellow students W. C. and E. R., he was a friend and a mentor.}
\affiliation{Department of Physics, Old Dominion University,
Norfolk, Virginia 23529, USA}

\author{William Clark}
\affiliation{Department of Physics, Old Dominion University,
Norfolk, Virginia 23529, USA}

\author{Antonia Seifert}
\affiliation{Institut für Theoretische Physik, Universit\"at Heidelberg, Philosophenweg 12, D-69120 Heidelberg, Germany}

\author{Alicia Mand}
\affiliation{Department of Physics, Old Dominion University,
Norfolk, Virginia 23529, USA}

\author{Emerson Rogers}
\affiliation{Department of Physics, Old Dominion University,
Norfolk, Virginia 23529, USA}

\author{Adam Lane}
\affiliation{Department of Physics, Old Dominion University,
Norfolk, Virginia 23529, USA}

\author{Alexandre Deur}
\affiliation{Department of Physics, University of Virginia,
Charlottesville, Virginia 22901, USA}
\affiliation{Department of Physics, Old Dominion University,
Norfolk, Virginia 23529, USA}

\author{Balsa Terzi{\'c}}
\affiliation{Department of Physics, Old Dominion University,
Norfolk, Virginia 23529, USA}

%\correspondingauthor{Bal{\V s}a Terzi{\'c}}
%\email{bterzic@odu.edu}

% Abstract (Do not insert blank lines, i.e. \\) 
\begin{abstract}
We examine the claimed observations of a gravitational external field effect (EFE) reported in Chae et al. 
We show that observations suggestive of the EFE can be interpreted without violating Einstein's equivalence principle, namely from known correlations between morphology, environment and dynamics of galaxies. 
While Chae et al's analysis provides a valuable attempt at a clear test of Modified Newtonian Dynamics, an evidently important topic, a re-analysis of the observational data does not permit us to confidently assess the presence of an EFE or to distinguish this interpretation from that proposed in this article.
\end{abstract}

% Keywords
%\keyword{Cosmology, General Relativity, dark energy, dark matter, MOND.} 

%%%%%%%%%%%%%%%%%%%%%%%%%%%%%%%%%%%%%%%%%%%%%%%%%

%%%%%%%%%%%%%%%%% BODY OF PAPER %%%%%%%%%%%%%%%%%%

\section{Introduction \label{sec:intro}}

One of the most pressing problems in contemporary science is the origin of an apparent substantial missing mass in large gravitational systems, such as galaxies or clusters of galaxies. The prevailing explanation for this phenomenon is the presence of non-baryonic, weakly-interacting matter known as dark matter. However, the consensus on dark matter's existence is not universal. This is partly due to the lack of detection of dark matter particles either directly~\citep{Kahlhoefer:2017dnp} or indirectly~\citep{Gaskins:2016cha}. Furthermore, advances in particle physics have ruled out the most natural extensions of the Standard Model of Particle Physics, which previously offered promising dark matter candidates \citep{Arcadi:2017kky}.
This has fueled alternative explanations for the missing mass problem. The most well-known of these alternatives is Modified Newtonian Dynamics (MOND) \citep{Milgrom:1983ca}. MOND has achieved notable success in economically describing observations at the galactic scale and even making predictions that have been subsequently validated, see \cite{Merritt:2020pwe,Banik_Zhao_2022} for review. One such prediction is the radial acceleration relation (RAR), a strict correlation between the measured accelerations experienced by baryonic matter within galaxies and the accelerations predicted by Newtonian gravity in the absence of dark matter. This strict correlation has been observed \citep{mcgaugh_2016, lelli:2017}. While a loose correlation may be expected within the context of dark matter and galaxy formation, they offer no uncontrived explanation for the strict correlation.
MOND is foremost a phenomenological theory. If its origin is fundamental rather than being an effective theory for phenomena arising from the complexity of galaxy dynamics and/or of General Relativity (GR), it would violate some of GR's tenets. Notably, the characteristic acceleration, denoted as $a_0$, should lead to an External Field Effect (EFE), which contradicts GR's strong equivalence principle.
The EFE would substantially weaken the self-gravity of dwarf galaxies in a cluster environment,
perhaps explaining why dwarf galaxies in the Fornax Cluster are much more susceptible to tides
than expected in LCDM \cite{Asencio_2022,Xu_2025}.

Recently, observations of EFE have been reported by \cite{chae_2020}. A refined analysis of the RAR data indicates that galaxies subjected to an external gravitational field, such as that emanating from neighboring galaxies, tend to display less missing mass compared to more isolated galaxies which, conversely, tend to have flatter rotation curves. This observation is naturally expected from MOND: given that the onset of MOND's regime starts below $a_0$, a constant external field added to the galactic field delays the transition to this regime. This, in turn, keeps the galaxy in the Newtonian regime longer, reducing their missing mass.

If these observations hold, it would mark the first experimental observation of a departure from GR and, equally significant, strongly challenge the dark matter hypothesis. Thus, it is imperative to study and exhaust alternative mundane explanations for the observation of \cite{chae_2020}. Its authors examine various such hypotheses and reject them. In this article, we examine the claim that the observations of \cite{chae_2020} evince the existence of an EFE. First, we investigate a possible explanation that has not been covered in \cite{chae_2020}, one that preserves GR and its strong equivalence principle while still obviating the need for dark matter. Specifically we show how correlations between galactic properties, rotation curves and environment may explain the \cite{chae_2020} observation. %Then, in Section~\ref{Bayesian analysis}, we follow the \cite{chae_2020} analysis to quantitatively compare how well MOND, with and without EFE, fits the observational data. 
We then conclude. Appendices provide technical details of the analyses in Section~\ref{GRSI-model} and a more quantitative analysis on how well MOND, with and without EFE, fits the observational data, based on the same MCMC approach taken by \cite{chae_2020}. 
In particular, we examine how the uncertainties associated with the \cite{chae_2020} analysis affect the claim of EFE observation.

\section{Modeling Radial Acceleration Relation with General Relativity Self-Interaction \label{GRSI-model}}
%{\color{red} ALEX: Outline of our argument and results.}
%{\color{blue}**AD: Our argument is very straightforward and natural. This advantage would be lost of we dilute it with the technical details. I will put the main steps of our work here and the result, referring to next sections for the technical details. We want to be as simple as possible while keeping the discussion honest (no crucial omission) and without detracting from understanding. Please noticed that I renamed the acceleration with labelled that are more suited in the context of the discussion: For ex. Chae use the label ``bar'' (for baryonic) because he discussed MOND against dark matter. We discuss GR-SI against MOND, so we don't need to contrast baryonic matter from dark matter.**\\
%
%-----------------------\\
%

The observation of \cite{chae_2020} can be naturally explained without jeopardizing the strong equivalence principle by considering two facts. The first is that rotation curves depend on the galaxy's distinct morphology. This is generally expected for any non-spherical extended system, regardless of whether Newtonian physics or GR is considered. If the system is massive enough, the self-interaction of the gravitational field in GR is expected to cause $V_\infty$ (the rotation speed at the outskirts of the galactic disk) to noticeably depend on the morphology of the galaxy, particularly on the relative size of the bulge versus that of the disk.
The second fact is the relationship between the environment and galactic morphology. It has been observed that the distribution of bulge types in the nearby universe clearly depends on the environment in which the galaxies reside \citep{fisher2011demographics}. This is expected in current hierarchical models of galaxy formation in which the bulges of disk galaxies grow from environmental disturbances, in particular mergers~\citep{baugh2006primer}, and is supported by the observation that the bulge-to-total mass ratio of nearby galaxies correlates with their number of satellite galaxies~\citep{javanmardi2020correlation}.
Taken together, the dependence of $V_\infty$ on morphology and the correlation of the latter with the environment could explain the observation by \cite{chae_2020} without violating the strong equivalence principle of GR. We will now quantitatively investigate this possibility.

To illustrate our model, we generated a set of 40,000 simulated disk galaxies with bulges sampling the correlations derived from observation, in the manner previously described 
in detail in \cite{deur_2020}.

We account for GR self-interaction (GR-SI) in a disk galaxy by modeling it following \cite{deur_2020}: the galaxy has two components, a disk and a bulge made solely of baryonic matter. The model does not require dark matter since for galaxies massive enough, field self-interaction constrains the gravitational field lines to remain within the disk, effectively confining gravity to the approximately two dimensions of the disk. This phenomenon, however, is suppressed in the bulge because its three-dimensional volume presents no particular direction or plane into which the field lines could collapse~\citep{Deur:2009ya}. This expectation has been verified by observing that in elliptical galaxies, the larger the ellipticity, the more the dynamics of baryonic matter departs from the Newtonian expectation~\citep{Deur:2013baa, Winters:2022tew}.

\begin{figure}
\includegraphics[width=0.48\textwidth, height=0.49\textwidth]{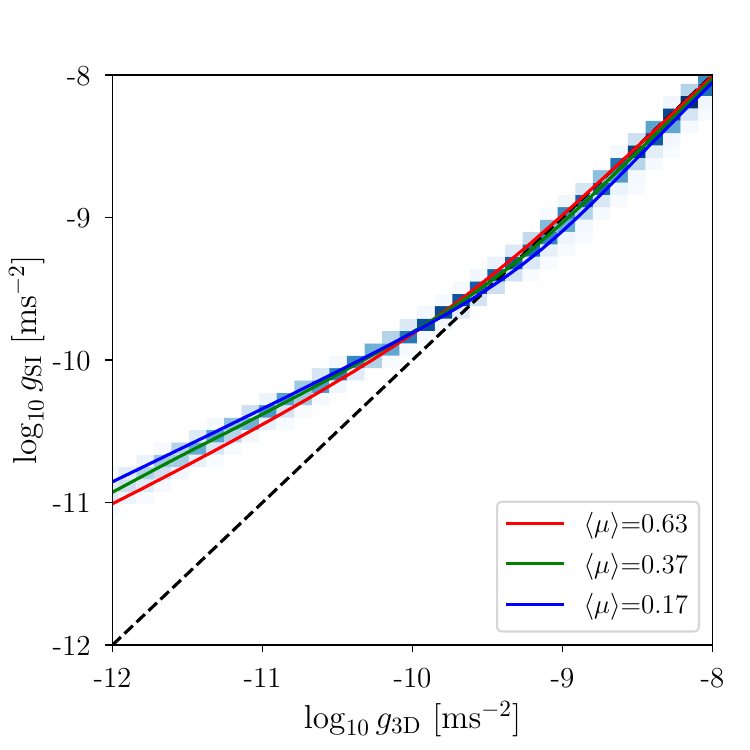}
\includegraphics[width=0.52\textwidth, height=0.52\textwidth]{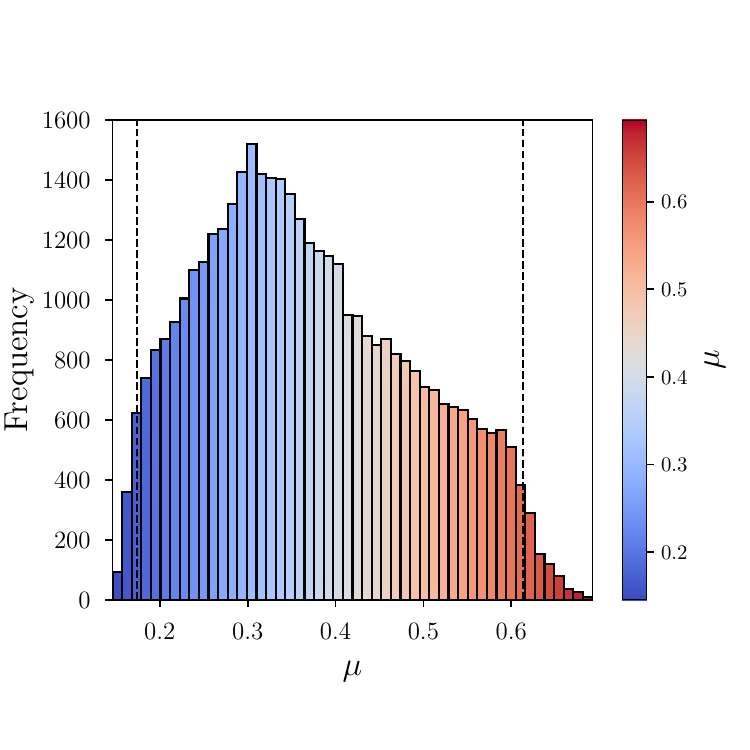}
\caption{Left: Radial acceleration relation for 40,000 simulated galaxies, color coded for bulge-to-total mass ratios $\mu$. Three different populations are shown, two of which are the edge cases of bulge-dominated (red line) and disk-dominated (blue line) galaxies (both subsamples contain 2\% of the total population), while the green line shows the remaining average population.
Right: Distribution of bulge-to-total mass ratio, $\mu$, for the same galaxies, also color-coded. The shape of the distribution comes from our enforcement of realistic combinations of galaxy parameters. Vertical dashed lines delineate the three populations. % \cite{deur_2020}. 
We use the 2D-gravity effective coupling value $\alpha=2.3\times 10^{-10}~{\rm m/s^2}$.
}
\label{fig:2pop}
\end{figure}

The model in \cite{deur_2020} entails that the
total acceleration within the galaxy ($g_{\rm SI}$) is the sum of the Newtonian acceleration ($g_{\rm 3D}$) from the buldge and that of gravity constrained to propagate in 2D ($g_{\rm 2D}$). We present an improved model: 
\begin{eqnarray}
 g_{\rm SI} (r)  &\equiv& g_{\rm 3D}(r) + g_{\rm 2D}(r) = G \frac{M_{\rm B}(r)}{r^2} + \sqrt{G \alpha} \frac{\sqrt{M_{\rm D}(r)}}{r},
 \label{GRSI_IF-1}
\end{eqnarray}
with $\sqrt{G\alpha}$ the effective coupling constant of gravity in two dimensions. 
$M_{\rm B}(r)$ and $M_{\rm D}(r)$ are the masses of the baryonic bulge and disk, respectively, 
contained within a sphere of radius $r$. We improved the model mainly with 
the handling of the factor $\alpha$ (discussed in detail in Appendix A), and so that it satisfies Tully-Fisher relation (detailed in Appendix B). The expression of the 2D-part relates to the effective potential emerging from the solution found by \cite{seifert2024new} for a system with cylindrical symmetry in GR, but is extrapolated to an extended source with given mass profile here.
It is worthwhile to emphasize that this and previous GR-based analyses of approximately 2D systems~\cite{Deur:2009ya, Deur:2017aas, Deur:2021ink} positions our approach strictly within the framework of GR and its approximations and the Standard Model of Particle Physics, in contrast to MOND and $\Lambda$CDM. 
% This approach is evaluated numerically constructing model galaxies by generating parameters such as half-light radius, disk scale 
% length, S{\'e}rsic parameter, total mass of disk, and total mass of bulge. 
% We then enforce known relations between these quantities in order to ensure 
% a realistic parameter space for individual galaxies. To determine $\rho_{\rm B}(r)$ 
% and $\rho_{\rm D}(r)$, we solve the Abel integral for the disk and bulge of individual galaxies.
% Normalizing these with $\rho_{\rm B0}$ and $\rho_{\rm D0}$ parameters obtained from the total 
% mass gives $M_{\rm B}(r)$ and $M_{\rm D}(r)$. Importantly, from this normalization follows that $\hat{m}_{\rm B}(r) \equiv M_{\rm B}(r) / M_{\rm B}^{\rm tot}$ and $\hat{m}_{\rm D}(r) \equiv M_{\rm D}(r) / M_{\rm D}^{\rm tot}$, where $M_{\rm B}^{\rm tot} \equiv \lim_{r\to\infty} M_{\rm B}(r)$ is the bulge mass and $M_{\rm D}^{\rm tot} \equiv \lim_{r\to\infty} M_{\rm D}(r)$ is the disk mass, are independent of the total mass.
The resulting $g_{\rm SI}$ is evaluated numerically (cf.~Appendix~B) and plotted versus $g_{\rm 3D}$ in Fig.~\ref{fig:2pop}. As the evaluation depends on the ratio of disk and bulge mass, or equivalently the bulge-to-total mass ratio $\mu \equiv M_{\rm B}^{\rm tot}/(M_{\rm B}^{\rm tot}+M_{\rm D}^{\rm tot})$, where $M_{\rm B}^{\rm tot}$ and $M_{\rm D}^{\rm tot}$ are total masses of the bulge and the disk, respectively, the distribution of this parameter in the numerical evaluation is plotted for reference.
Clearly, at low accelerations, galaxies with large bulges (red color in the figure)
are closer to the Newtonian expectation (diagonal line) than those with small bulge (blue color). 
Eq.~(\ref{GRSI_IF-1}) can be re-expressed as the ratio between the total acceleration 
to that from Newton's gravity:
\begin{equation} %\label{eq:ratio}
\frac{g_{\rm{SI}}}{g_{\rm 3D}} = 1 + \sqrt{\frac{\alpha}{G}} \frac{\sqrt{M_{\rm D}(r)}}{M_{\rm B}(r)}r,
\label{eq:if-si}
\end{equation}
which can then be compared to those proposed in \cite{mcgaugh_2016} and \cite{chae_2020} in the context of a violation of the strong equivalence principle, see Appendix B.

We now utilize the observed correlation reported in  \cite{javanmardi2020correlation}, wherein a correlation was found between the bulge-to-total mass ratio of galaxies with their number of satellite galaxies. This relationship is given by %Eq.~(\ref{JKR_1-1})
%**Jackson, I assumed that $B/T$ in your equations \ref{JKR_1}, \ref{JKR_2} is $M_{\rm B}/(M_{\rm D}+M_{\rm B})$, is that correct?** 
%
\begin{equation}
    \label{JKR_1-1}
    N^{200}=30.2(\pm 6.2)~\mu +1.7(\pm 1.3),
\end{equation}
where $N^{200}$ is the 
number of observed satellites within distance of 200 kpc from the main galaxy.
%where $M_V \leq -8.2$ mag. 
This correlation serves here as a measure of the density of the environment in which the main
galaxies reside, and therefore as a measure of the external gravitational field to which the 
galaxies are subjected.   
This approach differs from that of \cite{chae_2020} which estimated the external field within the standard $\Lambda$CDM context using an N-body simulation and the 2M++ survey \citep{lavaux20112m++} to populate observed SPARC galaxies. We find our approach to be appropriate for this study due to the difficulties $\Lambda$CDM has with estimating satellite populations \citep{Klypin:1999uc} and its challenges in explaining strong correlation between bulge mass and satellite populations \citep{javanmardi2019number}. The absence of accurate SPARC bulge-to-total ratios also makes a direct comparison to \cite{chae_2020} difficult. We also acknowledge that \cite{javanmardi2020correlation} did not aim to present an exact formula for $N$ vs.~$\mu$. Instead we use Eq.~(\ref{JKR_1-1}) for the purpose of better understanding the trend presented by the data. In doing so we note that the complex histories of individual galaxies play the most important role in determining morphology and corresponding isotropies. These complex histories present a problem when attempting to analyze individual galaxies. Therefore, analyzing individual rotation curves, while often useful, is not necessary in our context.

\begin{figure} 
\center
\includegraphics[width=0.49\textwidth]{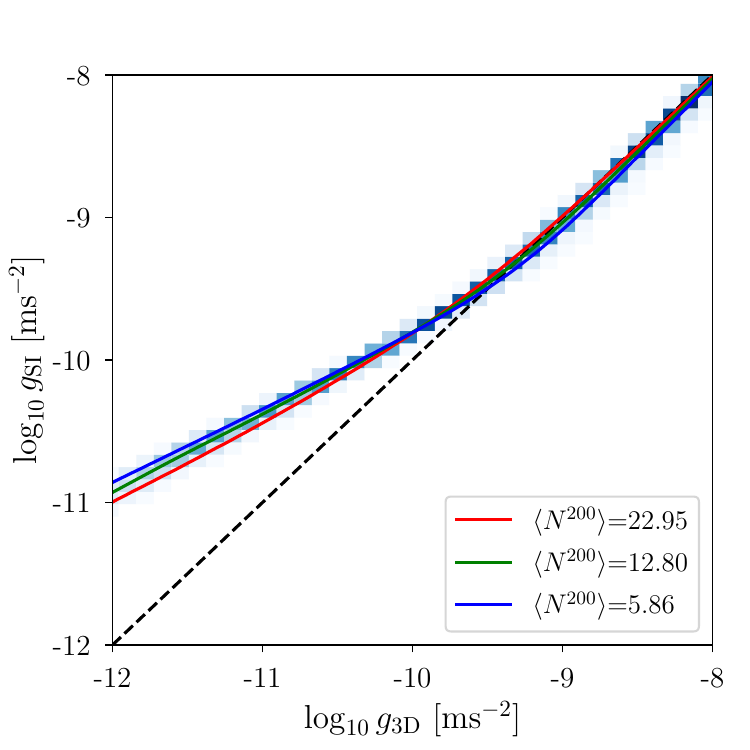}
\caption{Total acceleration $g_{\rm SI}$ versus that expected from the baryonic matter and Newtonian gravity, $g_{\rm 3D}$. The colors correspond to the average number of satellite  galaxies per bin within 200 kpc. The number of satellite galaxies were calculated by applying Eq.~(\ref{JKR_1-1}) to the bulge-to-total ratios of modeled galaxies. The similarity between Fig.~\ref{fig:g_vs_Nsat_1} and Fig.~\ref{fig:2pop} arises from both figures using the same 40,000 simulated galaxies and the strong relationship between the bulge-to-total mass ratio $\mu$ and the satellite galaxy population, as given by Eq.~(\ref{JKR_1-1}).}
\label{fig:g_vs_Nsat_1}
\end{figure}

Reformulating Eq.~(\ref{eq:if-si}) as done in Appendix B and combining Eqs.~(\ref{GRSI_nu_2}) and (\ref{JKR_1-1}) yields the 
observation that galaxies from a dense environment tend to be closer to the Newtonian expectation than isolated galaxies:
\begin{equation} \label{IF_N}
\log_{10} g_{\rm SI} = 
\frac{1}{2} \log_{10} g_{\rm 3D} + \frac{1}{2} \log_{10} \alpha 
+ \frac{1}{2} \log_{10} 
\left(\frac{30.2(\pm 6.2)}{N^{200}-1.7(\pm 1.3)} -1
\right).
\end{equation}
The third term on the right-hand side of  Eq.~(\ref{IF_N}) will result in the radial acceleration curve getting closer to the 45$^\circ$ diagonal as $N^{200}$ increases, as shown in Fig.~\ref{fig:g_vs_Nsat_1}.

%------------------------------------------------------------------------------------------------------
%------------------------------------------------------------------------------------------------------

%------------------------------------------------------------------------------------------------------
%------------------------------------------------------------------------------------------------------

Using this result we find we can thus account for the correlation observed by~\cite{chae_2020} without invoking an EFE. This is done by combining the verified GR-SI prediction that the departure from a Newtonian behavior of a massive system is suppressed by the degree of isotropy of the system~\citep{Deur:2013baa, Winters:2022tew}, with the observed correlations between the morphology of disk galaxies and satellite galaxy populations~\citep{fisher2011demographics, baugh2006primer, javanmardi2020correlation}. In short, if the dynamics of spiral galaxies are dependent on galactic morphology in the way predicted by GR-SI, and the morphology of galaxies are correlated with their number of satellite galaxies (and more generally their environment), then we expect to see a correlation between the dynamics of galaxies and their external fields. This expected correlation is in direct contrast to the explanation given by \cite{chae_2020} wherein external fields directly affect the dynamics of spiral galaxies via the suppression of the MOND effect.  
% Thus, observed correlations between the morphology of disk galaxies~\citep{fisher2011demographics, baugh2006primer, javanmardi2020correlation} and the verified GR-SI prediction that the departure from a Newtonian behavior of a massive system is suppressed by the degree of isotropy of the system~\citep{Deur:2013baa, Winters:2022tew}
% can account (without invoking an EFE) for the correlation observed by~\cite{chae_2020}.

%---------------------------------------------------

For a quantitative comparison, the MCMC analysis by \cite{chae_2020} should be repeated with our model. For technical details, see Appendix C. We find that the significance of the \cite{chae_2020} analysis is not as high as claimed. Re-conducting their analysis, we find considerably different values and large error margins. Within these errors, our results are consistent with the \cite{chae_2020} values but our analysis does not provide further insights aside from highlighting important instabilities in the MCMC analysis and possible insufficient convergence. This suggests that \cite{chae_2020} might not have found a global maximum of the likelihood in the parameter space. In particular, \cite{chae_2020} stop after a given number of iterations and do not demonstrate convergence of their MCMC analysis. To quantitatively determine which model is preferred would require a prohibitively expensive investigation of the parameter space to ensure proper convergence of the MCMC algorithm, which is beyond the scope of this paper. At this point, we can only state that the MCMC analysis by \cite{chae_2020} suffers from insufficient convergence, which, in addition to the alternative explanation presented here, further weakens claims for an EFE.

% \newpage
\section{Conclusion} \label{conclusion}
We examined the claim of observations of an external field effect (EFE) in gravitation reported in~\cite{chae_2020}, based on extensive data on disk galaxy rotation \citep{mcgaugh_2016}. First, we showed with an example that there exist other plausible explanations not requiring an EFE nor contrived models tailored to explain it away. 
To do so, we demonstrated that an apparent deviation from the Einstein's strong equivalence principle could be explained instead by examining the known correlations between galactic morphology, galactic environment (specifically, satellite galaxy populations), and galaxy dynamics. 
More precisely, it is expected from the non-linear character of general relativity that galaxies with larger bulges exhibit more Newtonian dynamics, just like rounder elliptical galaxies exhibit a smaller missing mass problem than flatter ones~\citep{Deur:2013baa, Winters:2022tew}.
This, together with the observation that galaxies with a higher number of satellites, which is indicative of a denser environment (and incidentally a greater external field), tend to have larger bulges~\citep{javanmardi2020correlation} explain the apparent EFE. 
Furthermore, while re-analyzing the data following the approach of \cite{chae_2020}, we found that, %the interpolation function accounting for EFE (viz, $\nu_{\rm C20}$ from \cite{chae_2020}) does not provide a significantly better description than the original interpolation function not including EFE (viz, $\nu_{\rm M16}$ from \cite{mcgaugh_2016}), nor that accounting for non-linearity of GR (viz, $\nu_{\rm SI}$ based on~\cite{deur_2020}). In other words, 
given the analysis method used by \cite{chae_2020} and here, 
observational data does not permit us
to confidently assess the the existence of an EFE or to distinguish between the possible various interpretations of these data.

\vskip10pt
%%%%%%%%%%%%%%%%%%%%%%%%%%%%%%%%%%%%%%%%%%
Conceptualization, C.S., A.D. and B.T.; methodology, C.S., A.D and B.T.; software, C.S., W.C., A.S., B.T.; validation, A.M., A.S., A.L., and E.R.; formal analysis, C.S., W.C., E.R., A.M., A.S., A.L. A.D. and B.T.; investigation, A.L.; resources, A.M, W.C.; data curation, C.S., A.M,; writing---original draft preparation, A.D., B.T.; writing---review and editing, A.D., B.T., W.C. and A.S.; visualization, C.S, W.C., A.S. and B.T.; supervision, A.D., B.T.; project administration, A.D., B.T.; funding acquisition, B.T. All authors have read and agreed to the published version of the manuscript.
\vskip5pt
This work was funded by U.S.~National Science Foundation grant No.~1847771 and by the Corey Sargent Memorial Scholarship in Physics at Old Dominion University.
\vskip5pt
All data used in this paper were collected from open sources and the references are provided.
\vskip5pt
%This publication is dedicated to the memory of Dr.~Corey Sargent, who tragically passed away on June 11, 2022, at the age of 25, and to his parents, Tracy Sargent and Gerard Carelli. At the time of his passing, Corey was developing the model at the heart of this article. For his fellow student W.~C.~and E.~R.~he was a friend and a mentor.
%\vskip5pt
The authors declare no conflict of interests.

%

%\section*{Acknowledgements}

%%%%%%%%%%%%%%%%%%%%%%%%%%%%%%%%%%%%%%%%%%%%%%%%%%



% \appendix

\section*{Appendix A. The Effective Coupling Constant $\alpha$} \label{sec:alpha-meaning}

The quantity $\sqrt{\alpha G}$ appearing in our model, Eq.~(\ref{GRSI_IF-1}), is the effective coupling of the gravitational force when its propagation is constrained within less than three 
dimensions. From dimensional consideration, a different coupling is required when a force propagates in space of dimension not equal to 3. This has been experimentally observed in the case of forces other than gravity. It is helpful here to use the strong nuclear force as an example which displays conspicuous field self-interaction effect.

The strong force potential between two static color charges separated by distance $r$ is accurately modeled by~\citep{Eichten:1974af, Eichten:1978tg, Eichten:2002qv}
\begin{equation}
V(r)=-4/3 \ \alpha_s/r + \sigma r, 
\end{equation}
where the 4/3 is a color combinatorial term coming from the fact that there are 3 types of color charge, 
$\alpha_s$ is the fundamental strong force coupling constant~\citep{Deur:2016tte, Deur:2023dzc, Gross_2022} and
$\sigma$ is the effective strong force coupling in 1 dimension, called the 
``string tension''. 
The term $-4\alpha_s/3r$ is the ``Coulomb/Newtonian'' part of the potential that dominates at 
short distances, while $\sigma r$ is the string/flux-tube part that dominates at long distance.
The latter term originates from the collapse of the force field lines into flux-tubes, and therefore 
is an {\it emergent} effective potential. In all, $\sigma$ is the coupling of this effective 1D force.

Likewise, $\sqrt{\alpha G}$ is the coupling of the effective force in 2D, in the case of a perfect disk galaxy which is massive enough to enable GR's non-linear regime. In the static cylindrical vacuum metric presented in \cite{seifert2024new}, an integration constant arises which, although \textit{a priori} without clear physical interpretation, compares to $\sqrt{\alpha G}$ in the effective potential. % it emerges as an integration constant without any other clear physical interpretation.

In the field line picture, it can be envisioned %It is possible 
that the galactic density is too weak to fully collapse the field lines into a 2D system. Therefore, the value of $\alpha$ will depend on the galaxy characteristics. Furthermore, the field line collapse occurs over a finite distance and, in particular, is delayed by the presence of the bulge. There is therefore a transition region where the value of $\alpha$ depends on the distance to the galactic center. This is not modeled in Eq.~(\ref{GRSI_IF-1}), which assumes only 3D and 2D regimes. From these considerations, it is not expected
that $\alpha$ is universal for all galaxies and it would need to be computed case
by case, which is not practical. Therefore, we treat here $\alpha$ as a free parameter of our model, albeit with a clear physical meaning.

\section*{Appendix B. Interpolation Function for General Relativity Self-Interaction}
\label{sec:int-func_GRSI}

The acceleration ratio $\nu$ for our model given in Eq.~(\ref{IF definition}) is
\begin{equation}
\nu_{\rm SI} (r) \equiv \frac{g_{\rm{SI}}(r)}{g_{\rm 3D}(r)} = 1 + \sqrt{\frac{\alpha}{G}} \frac{\sqrt{M_{\rm D}(r)}}{M_{\rm B}(r)}r\label{eq:if-si-1}.
\end{equation}
This approach is evaluated numerically by constructing model galaxies by generating parameters such as half-light radius, disk scale 
length, S{\'e}rsic parameter, total mass of the disk and the total mass of the bulge. 
We then enforce known relations between these quantities to ensure
a realistic parameter space for individual galaxies. To determine $\rho_{\rm B}(r)$ 
and $\rho_{\rm D}(r)$, we solve the Abel integral for the disk and bulge of individual galaxies.
Normalizing these with the parameters $\rho_{\rm B0}$ and $\rho_{\rm D0}$ obtained from the total
mass gives $M_{\rm B}(r)$ and $M_{\rm D}(r)$. Importantly, with this normalization we can rewrite 
%from this normalization follows that $\hat{m}_{\rm B}(r) \equiv M_{\rm B}(r) / M_{\rm B}^{\rm tot}$ and $\hat{m}_{\rm D}(r) \equiv M_{\rm D}(r) / M_{\rm D}^{\rm tot}$, where $M_{\rm B}^{\rm tot} \equiv \lim_{r\to\infty} M_{\rm B}(r)$ is the bulge mass and $M_{\rm D}^{\rm tot} \equiv \lim_{r\to\infty} M_{\rm D}(r)$ is the disk mass, are independent of the total mass.
% which, by the use of

\begin{align}
    M_{\rm B}(r) &= M_{\rm B}^{\rm tot} \hat{m}_{\rm B}(r) = \mu M^{\rm tot} \hat{m}_B(r),\\
    M_{\rm D}(r) &= M_{\rm D^{\rm tot}} \hat{m}_{\rm B}(r) = (1 - \mu) M^{\rm tot} \hat{m}_{\rm D}(r),
\end{align}
where  
$M^{\rm tot} \equiv M_{\rm B}^{\rm tot} + M_{\rm D}^{\rm tot}$ is the total baryonic mass.
The implicitly defined functions $\hat{m}_{\rm B}(r)$ and $\hat{m}_{\rm D}(r)$ are dimensionless mass profiles (in particular, independent of the total mass) which can be rescaled appropriately to obtain the mass profile of any synthetic galaxy with bulge and disk masses $M_{\rm B}^{\rm tot}, M_{\rm D}^{\rm tot}$, respectively. Using these definitions, we can reformulate Eq.~(\ref{eq:if-si-1}) as
\begin{align}
    \nu_{\rm SI} (r) = 1 + r\sqrt{\frac{\alpha}{G M^{\rm tot}}} \frac{\sqrt{(1 - \mu)\hat{m}_{\rm D}(r)}}{\mu \hat{m}_{\rm B}(r)}.\label{eq:if-si-mu}
\end{align}
For $\hat{m}_{\rm B}, \hat{m}_{\rm D}$ chosen from an appropriate galaxy model, this depends on the bulge-to-total-mass ratio $\mu$ only.

The ratio $\nu_{\rm SI}$ matches the Newtonian regime, $r \to 0$,
where $g_{\rm SI}(r)=g_{\rm 3D}(r)$ and $\lim_{r \to 0} \nu_{\rm SI}(r)=1$. The regime 
where SI is important, $r \to \infty$ where $g_{\rm SI}(r)=g_{\rm 2D}(r)$, yields
% It matches the Newtonian regime, $z \gg 1$ (or $r \to 0$)
% where $g_{\rm SI}(z)=g_{\rm 3D}(z)$ and $\lim_{z\to \infty} \nu_{\rm SI}(z)=1$ the regime 
% where SI is important, $z \ll 1$ (or $r \to \infty$), where $g_{\rm SI}(z)=g_{\rm 2D}(z)$ and
%
\begin{equation} \label{nu_SI_large_z}
\lim_{r\to \infty} \nu_{\rm SI}(r) = 
\sqrt{\frac{\alpha}{G}} \frac{\sqrt{M_{\rm D}^{\rm tot}}}{M_{\rm B}^{\rm tot}} r.
\end{equation}
From Eq.~(\ref{GRSI_IF-1}) we have
\begin{equation} \label{r_subs}
\lim_{r \to \infty} g_{\rm 3D}(r) = G \frac{M_{\rm B}^{\rm tot}}{r^2} 
\hskip10pt \Longrightarrow \hskip10pt
r = \sqrt{G M_{\rm B}^{\rm tot}/g_{\rm 3D}(r)}.
\end{equation}
Substituting Eq.~(\ref{r_subs}) in Eq.~(\ref{nu_SI_large_z}),
\begin{equation}
\lim_{r\to \infty} \nu_{\rm SI} (r) =
\sqrt{\frac{{\alpha M_{\rm D}^{\rm tot}}}{M_{\rm B}^{\rm tot} g_{\rm 3D}(r)}},
\end{equation}
and, multiplying it by $g_{\rm 3D}$, we obtain the asymptotic behavior of $g_{\rm SI}$: 
\begin{equation}
g_{\rm SI}(r) =
\sqrt{\frac{{ \alpha M_{\rm D}^{\rm tot}} g_{\rm 3D}(r)}{M_{\rm B}^{\rm tot}}},
\end{equation}
or, in terms of a $\log$-$\log$ plot of $g_{\rm SI}$ versus $g_{\rm 3D}$:
\begin{align} \label{GRSI_nu_2}
\log_{10} g_{\rm SI}(r) = 
\frac{1}{2} \log_{10} g_{\rm 3D}(r) + \frac{1}{2} \log_{10} \alpha 
+ \frac{1}{2} \log_{10} 
\left(\frac{M_{\rm D}^{\rm tot}}{M_{\rm B}^{\rm tot}}
\right).
\end{align}
Combining Eq.~(\ref{GRSI_nu_2}) with Eq.~(\ref{JKR_1-1}) yields Eq.~(\ref{IF_N}). The second term on the 
right-hand side of Eq.~({\ref{GRSI_nu_2}}) is model-dependent (related to the 
2D coupling $\alpha$), and the third term is a galaxy-dependent vertical shift 
(related to galaxy properties $M_{\rm B}$ and $M_{\rm D}$). 
Data~\citep{Sofue:2015tsa} shows that 
$M_{\rm B}^{\rm tot}$ and $M_{\rm D}^{\rm tot}$ are correlated via 
$M_{\rm D}^{\rm tot} = A (M_{\rm B}^{\rm tot})^{0.58(32)}$, where $A=e^{0.002(79)}$~\citep{deur_2020}. Incorporating this correlation into Eq.~(\ref{GRSI_nu_2}), we obtain
\begin{equation} \label{g_obs_asymp}
\log_{10} g_{\rm SI} (r)= \frac{1}{2} \log_{10} g_{\rm 3D}(r) + \frac{1}{2} \log_{10} \alpha + \frac{1}{2} \log_{10} A 
+ \frac{0.58(32)}{2} \log_{10} M_{\rm B}^{\rm tot}.
\end{equation}
From Eq.~(\ref{g_obs_asymp}), the larger $M_{\rm B}^{\rm tot}$, the closer the slope is to 1 (the diagonal in $g_{\rm obs}$ vs.~$g_{\rm 3D}$ plot), as evident from Fig.~\ref{fig:2pop}.

%
%---------------------------------------------------
%\begin{figure}
%\center
%\includegraphics[width=0.49\textwidth]{T3_bias_alpha=2.3e-10_Ngal=1000.pdf}
%\vspace{-0.4cm}
%\caption{The role of bias in $M_{\rm B}$ in the sample of 1000 galaxies.
%}
%\label{fig:updown}
%\end{figure}
%---------------------------------------------------

\begin{figure}
\center
\includegraphics[width=0.52\textwidth]{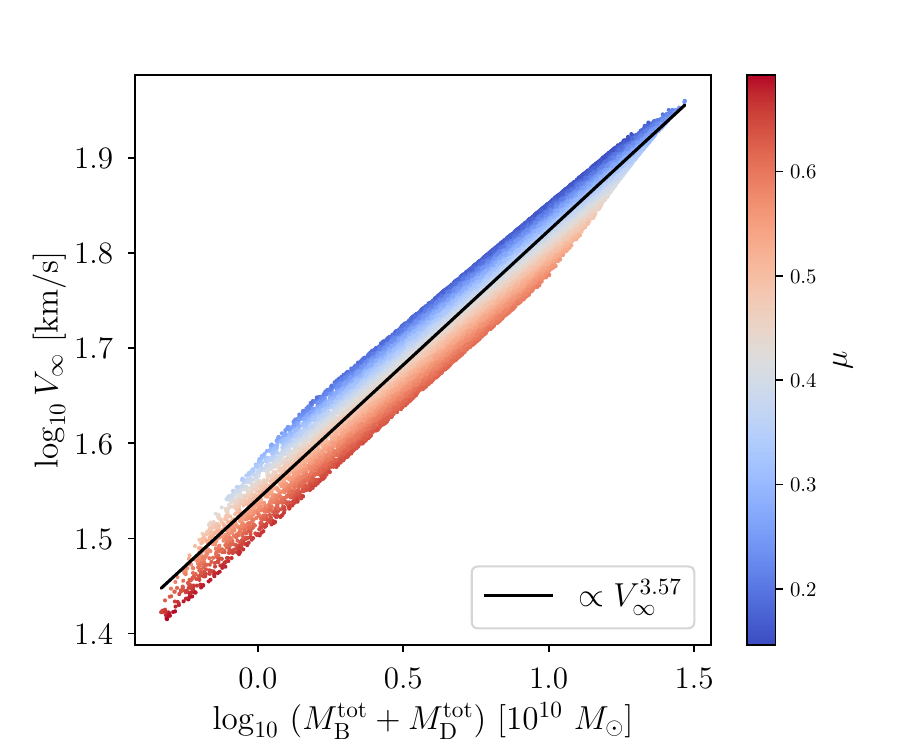}
\caption{Relationship between the total baryonic mass and the asymptotic velocity $V_\infty$ 
using the model in Eq.~(\ref{GRSI_IF-1}). A slope slightly smaller than 4 is consistent with 
baryonic Tully-Fisher relation. A value of $\alpha=2.3\times 10^{-10}~{\rm m/s^2}$ is used.}
\label{fig:BTFR}
\end{figure}

%From Eq.~(\ref{GRSI_IF}), the transition between the Newtonian and SI region occurs at $r=r_t$ when
%
%\begin{equation}
%\label{r_tran}
%G\frac{M_{\rm B}(r)}{r^2} = \sqrt{G\alpha} \frac{\sqrt{M_{\rm D}(r)}}{r}
%\hskip10pt \Longrightarrow \hskip10pt
%r_t = \frac{\sqrt{GM_{\rm D}(r_t)}}{\sqrt{\alpha}}.
%\end{equation}
%
%The expression above illustrates one of the fundamental differences between the MOND and
%GR SI: while they both feature two regimes---the inner Newtonian region and the outer
%MOND/SI regions---{\it the transition for MOND is fixed at $g^{\dagger}=a_0$}, while
%in GR SI, the constant $\alpha$ is fixed, and {\it the transition occurs at different radii
%$r_t$, depending on the galactic parameters}.
It can be shown from Eq.~(\ref{GRSI_IF-1}) that 
\begin{equation}
V_{\infty} = \lim_{r\to \infty} V(r) = \lim_{r\to \infty} \sqrt{g_{\rm 2D}(r)r} =
\left(\alpha G M_{\rm D}^{\rm tot}\right)^{1/4},     
\end{equation}
from which the relationship between the asymptotic velocity $V_{\rm \infty}$ and
the baryonic mass of the galaxy can be expressed as
\begin{equation}
M_{\rm D}^{\rm tot} = \frac{1}{\alpha G} V_{\infty}^4.    
\end{equation}
While this directly correlates the asymptotic velocity $V_{\infty}$ with the total baryonic
mass of the disk, the disk mass and the bulge mass being correlated, it is expected that
a relation between the total baryonic mass 
$M^{\rm tot} = M_{\rm B}^{\rm tot} + M_{\rm D}^{\rm tot}$  will also be strongly correlated
with the asymptotic velocity $V_{\infty}$. Figure \ref{fig:BTFR} shows 
this expectation, with a correlation slope
close to 4, consistent with the baryonic Tully-Fisher relation \cite{Lelli_2019}.

%\section{Fitting Interpolation Functions to Data}

%{\color{red}JACKSON: Please clean up this section and remove any redundancies from the main text.}

%---------------------------------------------------
%\begin{figure}
%\center
%\includegraphics[width=0.49\textwidth]{BTFR_alpha=1.2e-10_Ngal=1000.pdf}
%\vspace{-0.4cm}
%\caption{Relationship between the total baryonic mass and the {\color{magenta}\sout{terminal} asymptotic} velocity using
%the model in Eq.~\ref{GRSI_IF}. A slope
%slightly smaller than 4 is consistent with baryonic Tully-Fisher relation. A value of %$\alpha=1.2\times 10^{-10}~{\rm m/s^2}$ is used. {\color{red}JACKSON: Please replace this with %a figure which correspond to the previous figure, including hte gradient colors from blue to %red. No title.} }
%\label{fig:BTFR}
%\end{figure}

%---------------------------------------------------

\section*{Appendix C. Bayesian Analysis \label{Bayesian analysis}}

\begin{figure}
    \centering
    \includegraphics[width=\textwidth]{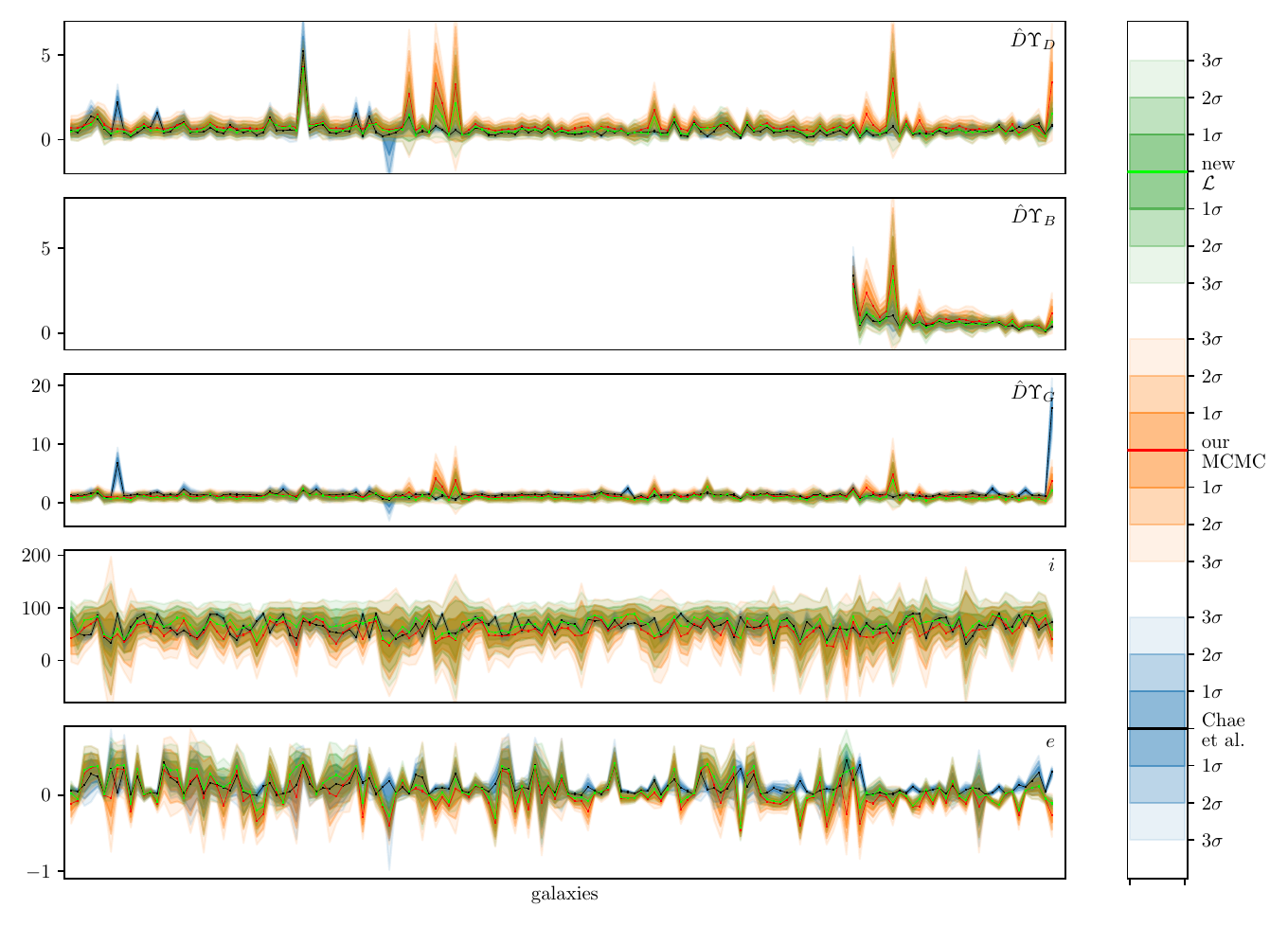}
    \caption{Results for the dimensionless parameters in the parameter vector and their error margins (1, 2 and 3$\sigma$) from the MCMC analysis by \cite{chae_2020} and in our evaluation, both using the $\nu_{\rm C20}$ model, Eq.~\eqref{eq:nu-model-c20}. The results from our analysis are given both for the likelihood build from $\chi^2$, Eq.~\eqref{eq:chi2}, (red) and $\hat{\chi}^2$, Eq.~\eqref{eq:chi2-new} (green). The estimated parameters are shown by the light green, red and black lines, respectively and the green, orange and blue regions give the respective error margins. In the first three panels, we show the products $\hat{D}\Upsilon_D$, $\hat{D}\Upsilon_B$, $\hat{D}\Upsilon_G$, respectively, as the mass-to-light rations $\Upsilon$ are degenerate with $\hat{D}$. Note that $\hat{D}\Upsilon_B$ can only be determined for galaxies with non-vanishing $V_B$. The two bottom panels give the results for the inclination $i$ and the external field parameter $e$.}
    \label{fig:mcmc-parameters}
\end{figure}

To compare our results to the observations, we use an approach similar to the one by \cite{chae_2020} formulated in terms of the interpolation function
\begin{equation}
\label{IF definition}
    \nu = \frac{g_{\rm model}}{g_{\rm observation}}.
\end{equation}
For our model presented in Section~\ref{GRSI-model}, this ratio is discussed in Appendix B. 
%\cite{chae_2020} give their interpolation function in Eq.~(5) {\color{magenta}**Do we mean Eq. 5 of their paper, not the one above? If so, we should be clearer about it. But is this sentence necessary anyway?}. 
The evaluation of the model is then done using Markov chain Monte Carlo (MCMC) simulations with the parameter vector $\vec \beta$ using the posterior probability
\begin{equation}
    p(\vec\beta) \propto \exp\left(-\frac{\chi^2}{2}\right) \prod\limits_k {\rm Pr}(\beta_k)\label{eq:prob},
\end{equation}
where ${\rm Pr}(\beta_k)$ are the prior probabilities and 
\begin{equation}
    \chi^2 = \sum\limits_{{\rm radial \ data \ points \ } R} \left(\frac{V_{\rm rot}(R) - V_{\rm model}(\vec\beta, R)}{\sigma_{V_{\rm rot}(R)}}\right)^2.\label{eq:chi2}
\end{equation}
This posterior probability follows in the work of \citet{chae_2020}. However, \citet{Chae_2022} later stated that \citet{chae_2020} used additional terms in the likelihood, namely
\begin{equation}
    \hat{\chi}^2 = \sum\limits_{{\rm radial \ data \ points \ } R} \left[\left(\frac{V_{\rm rot}(R) - V_{\rm model}(\vec\beta, R)}{\sigma_{V_{\rm rot}(R)}}\right)^2 + \ln \left(2 \pi \sigma_{V_{\rm rot}(R)}^2\right) \right] \equiv \chi^2 + \lambda.\label{eq:chi2-new}
\end{equation}
We examine the results for both choices of the posterior probability, but do not find any significant changes in our claims (Fig.~\ref{fig:mcmc-parameters}). In both cases, the
% The $\chi^2$ 
sum is calculated from the rotational velocity
\begin{align}
    V_{\rm rot} &= V_{\rm obs}\frac{\sin(i_{\rm obs})}{\sin(i)},\\
    \sigma_{V_{\rm rot}} &= \sigma_{V_{\rm obs}}\frac{\sin(i_{\rm obs})}{\sin(i)},
\end{align}
given by the observed values $V_{\rm obs}$ and their standard deviations $\sigma_{V_{\rm obs}}$ as well as the observed inclination $i_{\rm obs}$ and the theoretical inclination $i$ treated as a nuisance parameter in the MCMC. Furthermore, the baryonic matter velocity is calculated from the observed values for $V_{\rm B}(R)$ (bulge), $V_{\rm D}(R)$ (disk), $V_{\rm G}(R)$ (gas) \citep{SPARC:Lelli_2016}, the mass-to-light ratios $\Upsilon$ for each component and the fiducial distance $\hat D$, with the latter and the mass-to-light ratios being additional nuisance parameters to the MCMC. Here, $\Upsilon_G$ denotes the ratio of the total gas mass to the H\textsc{i} mass and is chosen in the same way as done by \citet{chae_2020}. In particular, for each galaxy we calculate the inverse ratio $X = \Upsilon_G^{-1}$ from \citet[Eq.~(4)]{chae_2020} and use it as a prior, allowing $\Upsilon_G$ to vary about $X^{-1}$ \citep[cf.][Table~1]{chae_2020}. The baryonic matter velocity is then given by 
\begin{align}
    V_{\rm bar}(R) &= \sqrt{\hat D \left(\Upsilon_D V_D^2(R) + \Upsilon_B V_B^2(R) + \Upsilon_G V_{\rm G}(R) \vert V_{\rm G}(R) \vert\right)},\label{eq:vbar}
\end{align}
following \cite{chae_2020}.
From this, the rotational velocity according to the model can be obtained from $\nu$, 
Eq.~(\ref{IF definition}),
\begin{equation}
    V_{\rm model}(\vec\beta, R) = \frac{V_{\rm model}(\vec\beta, R)}{V_{\rm bar}(\vec\beta, R)} V_{\rm bar}(\vec\beta, R) = \frac{\sqrt{Rg_{\rm model}(\vec\beta, R)}}{\sqrt{Rg_{\rm bar}(\vec\beta, R)}} V_{\rm bar}(\vec\beta, R) = \sqrt{\nu(\vec\beta, R)} V_{\rm bar}(\vec\beta, R).
\end{equation}
The parameter vector for the MCMC is given by $\vec\beta = (\Upsilon_{\rm D}, \Upsilon_{\rm B}, \Upsilon_{\rm G}, \hat D, i, [{\rm parameters \ for \ } \nu])$. The evaluation is done analogously to the analysis by \cite{chae_2020}, using the public code \texttt{emcee} \citep{Foreman_Mackey_2013} with $N_{\rm iteration} = 6000$, discarding the burn-in steps up to $N_{\rm iteration} = 500$ and thinning the rest by a factor of $50$. We choose $N_{\rm walkers} = 10,000$.

This evaluation can be performed for different interpolation functions $\nu$. Here, we examine the following functions: 
\begin{align}
    \nu_{\rm M16}(z) &= \frac{1}{1 - \exp({-\sqrt{z}})},\\
    \nu_{\rm C20}(z) &= \frac12 - \frac{e \left(1 + \frac{e}{2}\right)}{z(1+e)} + \sqrt{\left(\frac12 - \frac{e \left(1 + \frac{e}{2}\right)}{z(1+e)}\right)^2 + \frac{1 + e}{z}},\label{eq:nu-model-c20}\\
    \nu_{\rm SI} (R) &= \frac{g_{\rm SI}(R)}{g_{\rm 3D}(R)} = 1 + r\sqrt{\frac{\alpha}{G M^{\rm tot}}} \left\langle\frac{\sqrt{(1 - \mu)\hat{m}_{\rm D}(R)}}{\mu \hat{m}_{\rm B}(R)}\right\rangle_{\mu, \ \rm galaxy \ parameters},
\end{align}
where $e\equiv g_{\rm bar}/g_{\rm external}$ is the external field parameter, $M^{\rm tot}=M_{\rm B}^{\rm tot} + M_{\rm D}^{\rm tot}$ is the total mass, and $z = g_{\rm 3D}(R)/{g_\dagger}$ for constant 
$g_\dagger \equiv 1.2 \times 10^{-10} ~{\rm m} {\rm s}^{-2}$ as in \cite{chae_2020}.
The first of these, $\nu_{\rm M16}$, refers to the interpolation function presented in \cite{mcgaugh_2016}, the second, $\nu_{\rm C20}$, is the one from \cite{chae_2020}. Concerning $\nu_{\rm SI}$ for our model, we evaluate it based on Eq.~(\ref{eq:if-si-mu}). The normalized mass profiles $\hat{m}(r)$ are obtained as described in Appendix B, %Section~\ref{sec:gr-si-model}, 
choosing modeled galaxies with total mass matching the one of the observed galaxies. In particular, they are independent of the total mass, but we have to average over the bulge-to-total mass ratio $\mu$ and the galaxy parameters used in the galaxy model. The additional parameters $e$ for $\nu_{\rm C20}$ and $\alpha$ for $\nu_{\rm SI}$ are used as further parameters in the MCMC parameter vector, $\vec \beta$.

As all of these models depend on the mass-to-light ratios through $V_{\rm bar}$ only, the $\Upsilon_j$ cannot be determined if the respective velocity part $V_j$ vanishes for all radii. Furthermore, as the fiducial distance $\hat{D}$ enters the probability function through $V_{\rm bar}$ defined in Eq.~\eqref{eq:vbar} only, it is degenerate with the mass-to-light ratios but only the products $\hat{D}\Upsilon_D$, $\hat{D}\Upsilon_B$ and $\hat{D}\Upsilon_G$ are fixed. Comparing these factors and the remaining parameters resulting from the evaluation of $\nu_{\rm C20}$ to the results reported by \cite{chae_2020}, we find that they agree within the given error margins (Fig.~\ref{fig:mcmc-parameters}). 

Following \cite{chae_2020}, we use the Bayesian information criterion (${\rm BIC} \equiv -2 \ln L_{\rm max} + k \ln N$, with the maximum likelihood $L_{\rm max}$, the number of free parameters $k$, and the number of data points $N$) to compare the models.
From Eqs.~\eqref{eq:prob} and \eqref{eq:chi2-new} we find 
\begin{align}
    -2 \ln L_{\rm max} = \chi^2 + \lambda -2 \sum_k \ln \Pr(\beta_k).\label{eq:lmax}
\end{align}
Notably, the absolute value of the maximization likelihood is only fixed up to a constant by the MCMC parameters.
% Notably, the calculation of the ${\rm BIC}$ is based on the absolute values of the maximum likelihood, while the MCMC only relies on its maximisation property. \textcolor{red}{Thus, any constant factor in the likelihood will not influence the MCMC parameter values but only the ${\rm BIC}$. }
% In particular, using the absolute $\chi^2$ from Eq.~\eqref{eq:chi2} or the reduced $\Tilde{\chi}^2 = \frac{\chi^2}{ndf}$ with $ndf$ the number of degrees of freedom, does not change the MCMC results but is relevant for the value of the ${\rm BIC}$. \cite{chae_2020} do not state explicitly whether they use $\chi^2$ or $\Tilde{\chi}^2$ to determine the ${\rm BIC}$, but the values they give suggest that they consider the reduced $\Tilde{\chi}^2$, so we do this as well.
Furthermore, as the specific parameter values obtained from the MCMC differ from the results by \cite{chae_2020}, so does $L_{\rm max}$ and thus the ${\rm BIC}$. In particular, \cite{chae_2020} considered the so-called \textit{golden galaxies} (Table~\ref{tab:golden-gal}), two of which have very low $\Delta {\rm BIC}$ values according to \cite{chae_2020}, but not in our analysis. As we have conducted the same analysis as \citet{chae_2020} to the best of our knowledge, we conclude that the MCMC results are not stable enough to infer any results based on the $\Delta{\rm BIC}$ only. Investigating this criterion based on Eq.~\eqref{eq:lmax}, %We thus investigate the ${\rm BIC}$ based on Eq.~\eqref{eq:lmax} and $\Tilde{\chi}^2$, i.e., %by Eq.~(\ref{eq:prob}), we find $-2 \ln L_{\rm max} = \chi^2 -2 \sum_k \ln \Pr(\beta_k)$ and thus
\begin{equation}
    {\rm BIC} = \chi^2 + \lambda -2 \sum_k \ln \Pr(\beta_k) + k \ln N,
\end{equation}
we find that it is dominated by the $\chi^2$ and $\lambda$ terms. However, the $\lambda$ term does not differ much between the different models, as it depends only on one fitted parameter, $i$. The main contributions to any $\Delta{\rm BIC}$ value thus come from the $\chi^2$ term. In order to make any statements about the significance of the preference for any model, we analyze this term. In particular, we consider the reduced $\Tilde{\chi}^2 = \chi^2/ndf$ with $ndf$ the number of degrees of freedom. Note that any parameter vector that maximizes the likelihood in Eq.~\eqref{eq:prob}, also maximizes it for $\Tilde{\chi}^2$ instead of $\chi^2$ (or $\hat{\chi}^2/ndf$ instead of $\hat{\chi}^2$). Our conclusions are thus based
% which is dominated by $\Tilde{\chi}^2$ in our analysis. Thus, we focus 
on the difference in $\Tilde{\chi}^2$ given by $\Delta\Tilde{\chi}^2 = \Tilde{\chi}^2_{M16} - \Tilde{\chi}^2_{C20}$ %for our evaluation and determine 
with the corresponding uncertainty determined by error propagation based on the formula Eq.~(\ref{eq:chi2}) for $\Tilde{\chi}^2$.
This analysis yields no significant preferences of models for the golden galaxies (Table~\ref{tab:golden-gal}). We note that although we aimed to follow the approach described by \cite{chae_2020} to the best of our knowledge, the $\Delta {\rm BIC}$ and $\Delta \Tilde{\chi}^2$ values obtained from our analysis differ from the ones reported by \cite{chae_2020}. However, based on the error bars in our calculation of $\Delta \Tilde{\chi}^2$, this difference is not significant. 

To evaluate the preference of the models for the remaining galaxies, we use $\Delta\Tilde{\chi}^2$. For 135 out of 149 galaxies we find that this difference is within $1\sigma$, while two galaxies, UGC 3580 and NGC 2903, weakly favor the \cite{mcgaugh_2016} model and the other 12 are in favor of the model by \cite{chae_2020}. However, all of these preferences are still below the $2 \sigma$ significance. Comparing the models by \cite{mcgaugh_2016} and \cite{chae_2020} to GR-SI we find similar results. The MCMC based analysis of the interpolation function thus does not favor any of the three models. Importantly, we do not find any significant indication for an EFE. %Furthermore, analysing the error bars of $\chi^2$ we find that they mostly stem from the very broad error margins of the mass-to-light ratios fitted by the MCMC.

\begin{table}
    \centering
    % \begin{tabular}{cccc}
    %     \textbf{Galaxy} & \textbf{$\Delta {\rm BIC}$, \cite{chae_2020}} & \textbf{$\Delta {\rm BIC}$, our analysis} & \textbf{$\Delta\chi^2$, our analysis} \\ \hline
    %     NGC 5033 & $83.9$ & $113$ & $396 \pm 456$\\
    %     NGC 5055 & $144$ & $79503$ & $69716 \pm 69350$\\
    %     NGC 6674 & $< 10$ &  $14598$ & $5465 \pm 22437$\\
    %     NGC 1090 & $< 10$ & $3328$ & $2973 \pm 3017$\\ 
    % \end{tabular}
    % \begin{tabular}{cccccc}
    %     \textbf{Galaxy} & \textbf{$\Delta {\rm BIC}$, \cite{chae_2020}} & \textbf{$\Delta {\rm BIC}$, our analysis} & \textbf{$\Tilde{\chi}^2_{\rm M16}$, our analysis} & \textbf{$\Tilde{\chi}^2_{\rm C20}$, our analysis} & \textbf{$\Delta\Tilde{\chi}^2$, our analysis} \\ \hline
    %     NGC 5033 & $83.9$ & $19$ & $31 \pm 27$ & $8.0 \pm 1.4$ & $23 \pm 27$\\
    %     NGC 5055 & $144$ & $2811$ & $3332 \pm 2863$ & $507 \pm 416$ & $2824 \pm 2893$\\
    %     NGC 6674 & $< 10$ &  $-38$ & $1503 \pm 1556$ & $1519 \pm 2308$ & $-16 \pm 2783$\\
    %     NGC 1090 & $< 10$ & $143$ & $178 \pm 147$ & $30 \pm 34$ & $148 \pm 151$\\ 
    % \end{tabular}
    \begin{tabular}{cccccc}
        \textbf{Galaxy} & \textbf{$\Delta {\rm BIC}$, \cite{chae_2020}} & \textbf{$\Delta {\rm BIC}$, our} & \textbf{$\Tilde{\chi}^2_{\rm M16}$, our} & \textbf{$\Tilde{\chi}^2_{\rm C20}$, our} & \textbf{$\Delta\Tilde{\chi}^2$, our} \\ \hline
        NGC 5033 & $83.9$ & 
    $3$ & $31 \pm 27$ & $8.0 \pm 1.4$ & $23 \pm 27$\\
        NGC 5055 & $144$ & 
    $66803$ & $3332 \pm 2863$ & $507 \pm 416$ & $2824 \pm 2893$\\
        NGC 6674 & $< 10$ &  
    $4994$ & $1503 \pm 1556$ & $1519 \pm 2308$ & $-16 \pm 2783$\\
        NGC 1090 & $< 10$ & 
    $2830$ & $178 \pm 147$ & $30 \pm 34$ & $148 \pm 151$\\ 
    \end{tabular}
    \caption{Evaluation based on $\Delta {\rm BIC}$ for the analyses of $\nu_{\rm M16}$ and $\nu_{\rm C20}$ for the golden galaxies investigated by \cite{chae_2020} compared to our analysis. Positive values indicate preference of the \cite{chae_2020} model with EFE over the one by \cite{mcgaugh_2016}.
    The first column provides the values reported by \cite{chae_2020} and in the second, we present the values obtained from our MCMC analysis of the $\nu_{\rm M16}$ and $\nu_{\rm C20}$ interpolation functions. For investigating the significance of the model preference, we provide the $\Tilde{\chi}^2$ values for the $\nu_{\rm M16}$ (third column) and $\nu_{\rm C20}$ (fourth column) interpolation functions as well as their difference $\Delta \Tilde{\chi}^2$ with the respective error margins in addition to the ${\rm BIC}$. Note that the $\Delta{\rm BIC}$ values are quoted based on $\hat{\chi}^2 = \chi^2 + \lambda$ from Eq.~\eqref{eq:chi2-new}, whereas the three right columns give the difference in \textit{reduced} $\Tilde{\chi}^2 = \chi^2/ndf$, making it less obvious that the ${\rm BIC}$ is indeed dominated by $\chi^2$.}
    \label{tab:golden-gal}
\end{table}

%{\color{green} We find it necessary to briefly discuss %the significance of the "Golden galaxy" argument %presented in \cite{chae_2020}. In the golden galaxy %argument the authors analyzed 4 SPARC galaxies with %similar morphology's. These galaxies were selected %because of the strength of their estimated external %fields. Two of these galaxies (NGC 5033 and NGC 5055) %had among the strongest external fields calculated while %the other two (NGC 6674 and NGC 1090) had some of the %smallest. The authors found that the two galaxies with %the largest estimated external fields showed the largest %suppression of the MOND effect while the other two %showed less MOND suppression. Due to the sensitivity of %GR-SI on disk and bulge mass densities ( in particular %bulge-to-total ratios) one can not test our model on the %basis of such a sample of four galaxies, rather a test %of galaxies with similar bulge-to-total ratios (and %differing external gravitational fields) would be more %useful in the context of GR-SI.}

%\iffalse
%\begin{figure}
%\label{fig:g_vs_Nsat}
%\centering
%\includegraphics[width=0.49\textwidth]{g_vs_Nsat.png}
%\includegraphics[width=0.49\textwidth]{chae.png}
%\vspace{-0.4cm}
%\caption{Plot showing that as the number of near neighbors increases galaxies become more Newtonian. This come directly from the %linear relationship between $N$ and $B/T$ found in Eq.~(\ref{JKR_1}).}
%\end{figure}
%\fi

\bibliography{Sargent_etal_2025}{}
%\bibliographystyle{aasjournal}

%%%%%%%%%%%%%%%%%%%%%%%%%%%%%%%%%%%%%%%%%%%%%%%%%%

% Don't change these lines
%\bsp	% typesetting comment
%\label{lastpage}

\end{document}